\documentclass[journal,twoside,web]{ieeecolor}
\usepackage{tmi}
\usepackage{cite}
\usepackage{amsmath,amssymb,amsfonts}
\usepackage{algorithmic}
\usepackage{tabularx}
\usepackage{graphicx}
\usepackage{textcomp}
\def\BibTeX{{\rm B\kern-.05em{\sc i\kern-.025em b}\kern-.08em
		T\kern-.1667em\lower.7ex\hbox{E}\kern-.125emX}}
\begin{document}
	\title{An explainability framework for cortical surface-based deep learning}
	\author{Fernanda L. Ribeiro, Steffen Bollmann, Ross Cunnington, and Alexander M. Puckett
		\thanks{This work was supported by the Australian Research Council (DE180100433). Data were provided by the Human Connectome Project, WU-Minn Consortium (Principal Investigators: David Van Essen and Kamil Ugurbil; 1U54MH091657) funded by the 16 NIH Institutes and Centers that support the NIH Blueprint for Neuroscience Research; and by the McDonnell Center for Systems Neuroscience at Washington University. The authors acknowledge Jake Carroll and Irek Porebski for their great support with the high-performance computing facilities of the Research Computing Centre at the University of Queensland.}
		\thanks{F. L. Ribeiro is with the School of Psychology and Queensland Brain Institute at the University of Queensland, Brisbane QLD 4072, Australia (e-mail: fernanda.ribeiro@uq.edu.au).} 
		\thanks{S. Bollmann is with the School of Information Technology and Electrical Engineering at the University of Queensland, Brisbane QLD 4072, Australia (e-mail: s.bollmann@uq.edu.au).} 
		\thanks{R. Cunnington is with the School of Psychology at the University of Queensland, Brisbane QLD 4072, Australia (e-mail: r.cunnington@uq.edu.au).} 
		\thanks{A. M. Puckett is with the School of Psychology and Queensland Brain Institute at the University of Queensland, Brisbane QLD 4072, Australia (e-mail: a.puckett@uq.edu.au).}}

	\maketitle
	\IEEEpeerreviewmaketitle
	
	\begin{abstract}
		The emergence of explainability methods has enabled a better comprehension of how deep neural networks operate through concepts that are easily understood and implemented by the end user. While most explainability methods have been designed for traditional deep learning, some have been further developed for geometric deep learning, in which data are predominantly represented as graphs. These representations are regularly derived from medical imaging data, particularly in the field of neuroimaging, in which graphs are used to represent brain structural and functional wiring patterns (brain connectomes) and cortical surface models are used to represent the anatomical structure of the brain. Although explainability techniques have been developed for identifying important vertices (brain areas) and features for graph classification, these methods are still lacking for more complex tasks, such as surface-based modality transfer (or vertex-wise regression). Here, we address the need for surface-based explainability approaches by developing a framework for cortical surface-based deep learning, providing a transparent system for modality transfer tasks. First, we adapted a perturbation-based approach for use with surface data. Then, we applied our perturbation-based method to investigate the key features and vertices used by a geometric deep learning model developed to predict brain function from anatomy directly on a cortical surface model. We show that our explainability framework is not only able to identify important features and their spatial location but that it is also reliable and valid.
	\end{abstract}
	
	
	\section{Introduction}
	\label{sec:introduction}
	\IEEEPARstart{T}{here} is no question about the utility of deep learning, yet it has long been criticized for its opacity. This opacity arises from the complexity of deep learning models, which are often marked by millions of parameters. The “black box” nature of these models has limited their application to the field of medical imaging, given that model explainability is critical for human verification that a system’s reasoning is fair and acceptable – particularly for clinical use\cite{He2019}. For instance, algorithmic bias with respect to protected attributes, such as sex and race, can potentiate discriminatory practices\cite{Char2018,Larrazabal2020,Seyyed-Kalantari2021}. However, the emergence of explainability methods\cite{Ras2020} has enabled a better comprehension of how these models operate through concepts that are easily understood and analyzed by the end user\cite{Hagras2018} – for example, by determining the key features driving a specific class prediction. 
	
	While most explainability methods have been designed for traditional deep learning, i.e., deep learning applied to Euclidean data, some have been further developed for geometric deep learning\cite{Baldassarre2019,Huang2020,Ying2019}, in which data are predominantly represented as graphs. Graphs are non-Euclidean, mathematical representations of individual entities (nodes/vertices) and their associations (connections/edges). These representations are regularly derived from medical imaging data, particularly in the field of neuroimaging. Two commonly used examples are (1) “connectomes” which represent functional and anatomical associations between brain regions and (2) cortical surface models which are sparse graphs – composed of vertices and edges that form triangular faces folded up in 3D space – that are used to represent the anatomical structure of the brain. 
	
	Connectomes are often used to investigate interactions between neurons, brain regions, and systems that culminate in complex behaviors\cite{Finn2015,Shen2017}, and their disruptions are associated with neurological and neuropsychiatric disorders\cite{Gratton2019,Hall2019,VandenHeuvel2019}. On the other hand, the utility of cortical surface models includes, but is not limited to: determining the topographical organization of sensory cortices\cite{DaCosta2015,Dougherty2003,Hinds2009,SanchezPanchuelo2018}; establishing the hierarchical organization of brain activity regarding its topology and how it underpins complex behaviors\cite{Margulies2016}; investigating developmental changes in cortical structure and function\cite{Sydnor2021,Zhao2021}; and shape analysis for disease detection\cite{Sarasua2021}. Despite equal relevance, geometric deep learning has predominantly been applied for the classification of connectomes with respect to gender\cite{Xing2021}, disease\cite{Li2021,Parisot2018,Xing2021}, and brain state\cite{Li2021,Zhang2021} (for more see \cite{Bessadok2021}). From this previous work, explainability techniques have been developed for distilling connectomes into important nodes and features for graph classification and, to a limited extent, regression tasks\cite{Li2021}. These methods, however, are still lacking for more complex tasks, such as surface-based modality transfer or vertex-wise regression\cite{Ribeiro2021,Zhao2019,Zhao2021}.
	
	We address the need for surface-based explainability approaches by developing a framework for cortical surface-based deep learning\cite{Parvathaneni2019,Ribeiro2021,Seong2018,Zhao2019,Zhao2021}, providing a transparent system for modality transfer tasks. To this end, we developed a perturbation-based approach capable of identifying important features as well as their spatial location on the anatomical surface model. Among different categories of foundational explainability approaches\cite{Ras2020}, perturbation-based methods aim to extract meaning from a deep learning model by visualizing input characteristics that strongly influence the model’s prediction, typically through saliency maps\cite{Ivanovs2021}. Although originally developed for application to Euclidean data (e.g., to identify “salient” pixels in 2D images), this approach is particularly well-suited for applications to cortical surface models given the meaningful spatial relationship among surface vertices. 
	
	The saliency maps produced by a perturbation-based method express the degree of importance of different spatial units in the input space (i.e., pixels in 2D images or vertices in surfaces) to the model’s prediction. For instance, saliency maps can highlight the importance of different pixels for the correct class probability in image classification problems. Zeiler and Fergus (2014), for example, developed a perturbation-based approach in which they systematically occluded different regions of an input image by sweeping a gray square over the image and monitoring changes in the output of the model\cite{Zeiler2014}. This iterative process results in a map of correct class probability as a function of the spatial location of the gray square, which is referred to as a saliency map. Correct class probability drops where important pixels are occluded, in contrast to unimportant pixels. Therefore, the saliency map highlights the spatial location of key pixels to predict the correct image class. This approach is particularly interesting for medical imaging since it allows for ‘knockout’ experiments, in which localized areas can be modified or occluded in a hypothesis-driven manner. Moreover, this methodology is model agnostic; it applies to any machine learning model regardless of its architecture. Therefore, perturbation-based methods represent an intuitive way of highlighting the input-output relationship learned by the model. 
	
\begin{figure*}[ht]
		\centerline{\includegraphics[width=2\columnwidth]{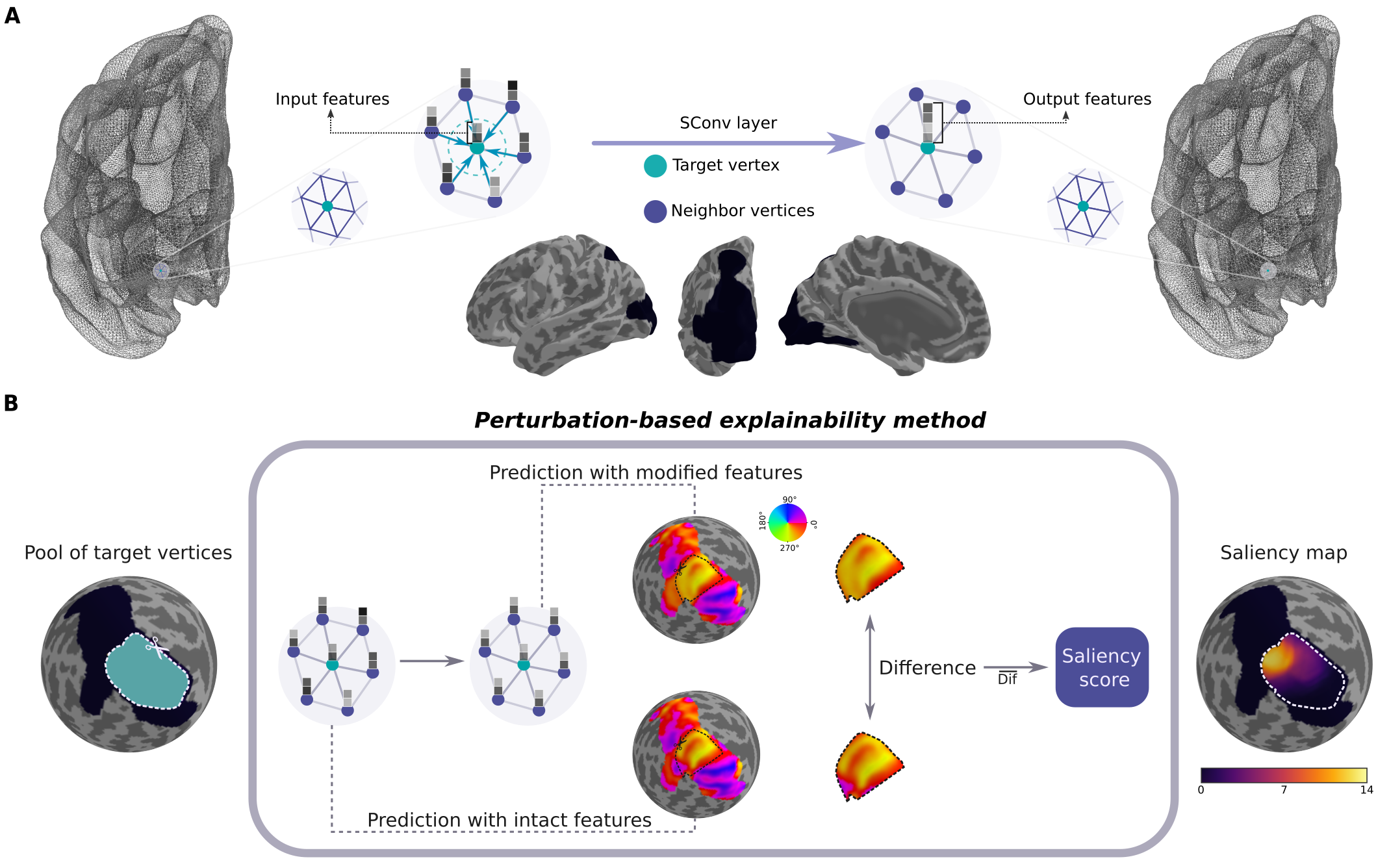}}
		\caption{\textbf{Perturbation-based explainability method for modality transfer tasks. A,} Convolutional layers in our geometric deep learning model aggregate vertex features in local neighborhoods weighted by learnable parameters of a continuous kernel function\cite{Fey2018}. This process generates new attributes for each target vertex at a time. Vertices’ features varied across participants reflecting their anatomical structure. On the other hand, the cortical surface space was the same for all individuals, allowing for vertex correspondence across individuals. We trained the model on a subset of vertices, illustrated by the black area in the middle images. \textbf{B,} We iteratively selected a target vertex and its N-hop neighborhood, replaced their features (myelin and curvature) with constants, and generated new polar angle maps for all participants in the test dataset with the previously trained model. The difference between polar angle predictions with intact versus modified input was determined in a vertex-wise manner and averaged across vertices in the dorsal portion of early visual cortex and participants in the test dataset, resulting in a single value per vertex – the saliency score. The resulting saliency map shows vertices’ saliency scores as a function of their spatial location on the surface.}
		\label{fig1}
	\end{figure*}

	Here, we adapted the perturbation-based method of Zeiler and Fergus (2014) for use with surface data to establish an explainability framework for surface-based deep learning (Figure \ref{fig1}). Like patches of 2D images (Euclidean data), neighborhoods of vertices in surface representations (non-Euclidean data) highlight spatially contiguous units of data. Consequently, a perturbation-based method can be similarly applied to surface data by systematically occluding input features that fall within a target vertex’s neighborhood and determining the difference in the model’s predictions (intact versus modified input). Once established, we applied our perturbation-based method to investigate the key features and vertices used by a geometric deep learning model previously developed to predict brain function from anatomy directly on a cortical surface model\cite{Ribeiro2021}. We show that our explainability framework is not only able to identify important features and their spatial location but that it is also reliable and valid.
	
	\section{Related work}
	\subsection{Retinotopic mapping with geometric deep learning }
	The spatial organization of sensory input reaching the retina is maintained in visual areas, although in a distorted manner\cite{Cohen2011,Cowey1974}, characterizing the topographical organization of the visual cortex. In other words, adjacent neurons in these retinotopically organized visual areas map adjacent locations in the visual field. Moreover, sensory input is processed through multiple visual areas, with later areas capturing increasingly more complex features of visual scenes\cite{Grill-Spector2004}. So, the visual cortex is retinotopically as well as hierarchically organized. These two properties are intertwined anatomically, as transitions in retinotopic maps delineate boundaries of visual areas throughout the visual cortex hierarchy\cite{Wandell2010}. 
	
	Although empirical retinotopic mapping has been the primary means of delineating precise visual area boundaries in individuals, retinotopic map models have shown great potential in accounting for individual variability\cite{Benson2012,Benson2018a,Ribeiro2021}. In particular, inter-individual variability in retinotopic maps has been directly associated with variability in anatomical properties, such as brain size\cite{Dougherty2003}, cortical folding patterns\cite{Benson2012,Hinds2009,Rajimehr2009}, and the degree of myelination\cite{Abdollahi2014}. Therefore, models of retinotopy have been proposed to leverage this structure-function relationship in the visual cortex. As follows, the cortical shape has been a valuable predictor of retinotopic organization in early visual cortex\cite{Benson2012,Hinds2009,Wang2015}. Nonetheless, these approaches rely on templates that are warped to an individual’s cortical anatomy and can only account for gross individual differences in the functional organization of these areas\cite{Benson2012}. Thus, we recently proposed a geometric deep learning model\cite{Ribeiro2021} to represent this intricate structure-function relationship without enforcing a spatially consistent mapping (i.e., a template), such that more idiosyncratic maps could be predicted. To do so, we trained models to predict functional maps (polar angle maps) from anatomical information (curvature and myelin maps) on a cortical surface model of the human brain (Figure \ref{fig1}A). We found that our model was not only able to predict the functional organization of human visual cortex from anatomical features, but it was also able to predict nuanced variations in polar angle maps across individuals\cite{Ribeiro2021}.
	
	The unprecedented details in predicted retinotopic maps brought into question the characteristics of the input data relevant for this individual variability. Specifically, our findings corroborated previous reports suggesting that the dorsal portion of early visual cortex is highly variable across individuals\cite{Allen2021,Arcaro2015a,Benson2018a,VanEssen2018}. Indeed, our model was able to predict these unique variations suggesting that the idiosyncrasy noticed in this region likely reflects real structure-related variation worthy of further investigation. This further investigation was one of the motivating factors for developing our explainability framework. In fact, we were interested in determining a framework to answer where and what were the key features and vertices for determining these idiosyncratic patterns. Hence, we show how this explainability technique elucidates the input-output relationship learned by a geometric deep learning model of retinotopy, which can be further exploited to understand the link between brain structure and function in medical imaging data.

	\subsection{Explainable geometric deep learning }
	Although still a budding area of artificial intelligence (AI) research, explainable AI already has a considerable body of work spanning three main categories: visualization methods, model distillation, and intrinsic methods\cite{Ras2020}. Explainability work for geometric deep learning has already touched upon these major categories. For example, Baldassarre and Azizpour (2019) have explored the ability of backpropagation-based methods, which are situated within the visualization category since they rely on explanations provided by saliency maps\cite{Ras2020}, to generate valid explanations for two graph-based tasks: node classification and graph regression. In both cases, Layer-wise Relevance Propagation\cite{Bach2015} has generated explanations consistent with ground truth knowledge\cite{Baldassarre2019}. GNNExplainer\cite{Ying2019}, on the other hand, combines different facets of visualization and model distillation methods. It takes as input a trained model and its predictions, and it learns a small subgraph of maximum mutual information with the model’s predictions. Hence, it essentially distills the most relevant subgraph from the input graph. Noteworthy, this method is model agnostic and is suitable for node classification, link prediction, and graph classification. GraphLIME\cite{Huang2020}, a model distillation method, learns a local nonlinear interpretable model for a node’s predicted class given its N-hop neighborhood. Despite also being model agnostic, it has only been validated for node classification tasks. Finally, BrainGNN\cite{Li2021} is an intrinsically interpretable model built for brain graph classification; by design, it keeps the most salient brain regions, which are leveraged for explanations. Crucially, salient regions in a brain state classification task corresponded with biomarkers reported in the literature\cite{Li2021}. Overall, none of the abovementioned methods is suited for generating explanations for node regression tasks (or vertex-wise regression). 
	
	\subsection{Contributions}
	Here, we developed a perturbation-based explainability method for vertex-wise regression tasks. We build upon visualization methods to develop a data-centric explainability framework for cortical surface-based deep learning. That is, we learn about how a geometric deep learning model of retinotopy works through systematic manipulations of the input data and measurement of changes in the model’s output. We adapted a perturbation-based approach\cite{Zeiler2014} to surface data, such that the perturbation kernel takes into consideration a target vertex’s N-hop neighborhood. Furthermore, we propose two strategies to validate our explainability method: (\textbf{1}) we train new models with relevant features replaced with a constant value, and (\textbf{2}) we mask important vertices from the loss function for semi-supervised training. Although we exemplify the use of the proposed approach to elucidate a model of retinotopy, the method is not limited to this case study; it is a general framework for surface-based deep learning. Therefore, the main contributions of our work are:

	\begin{itemize}
		\item	A novel formulation of a perturbation-based explainability method for surface-based deep learning.
		
		\item	An explainability method suitable for investigating the input-output relationship in modality transfer tasks.
		
		\item	A new framework to determine reliability and validity of explainability methods in the absence of ground truth knowledge.
	\end{itemize}

	\section{Methods}
	
	\subsection{Dataset}
	We used the Human Connectome Project (HCP) 7T Retinotopy dataset\cite{Benson2018b} for modeling and validation of our perturbation-based explainability approach. This dataset consists of high-resolution functional retinotopic mapping and structural data from 181 participants (109 females, age 22-35) with normal or corrected-to-normal visual acuity. Participant recruitment and data collection were led by Washington University and the University of Minnesota. The Institutional Review Board (IRB) at Washington University approved all experimental procedures (IRB number 201204036; “Mapping the Human Connectome: Structure, Function, and Heritability”), and all participants provided written informed consent before data collection\cite{VanEssen2013}. Additionally, the acquisition protocol has been described in previous work\cite{Benson2018b,VanEssen2013}. 
	
	Structural data were acquired at 0.7 mm isotropic resolution in a customized Siemens 3T Connectome scanner\cite{VanEssen2013}. Cortical surfaces were reconstructed from T1w structural images using FreeSurfer and aligned to the 32k fs\_LR standard surface space. This standard 32k fs\_LR cortical surface consists of 32,492 vertices sparsely connected, forming triangular faces. Curvature maps were generated from native surfaces and resampled to the 32k fs\_LR standard surface space. Similarly, myelin maps were determined by the ratio of T1w/T2w images\cite{Glasser2011}, followed by normalization to account for B1+ transmit effects and resampling to the 32k fs\_LR standard surface\cite{Glasser2013}. 
	
	Functional retinotopic mapping data were acquired using a Siemens 7T Magnetom scanner at a resolution of 1.6mm isotropic and 1s TR. Data were preprocessed following the HCP pipeline, which included correction for head motion and EPI spatial distortion, alignment of the fMRI data with the HCP standard surface space, and denoising for spatially specific structured noise. Retinotopic mapping stimuli comprised rotating wedges, expanding and contracting rings, and bars of different orientations moving across the visual field in different directions. Appropriate analysis of neural responses elicited by the known spatiotemporal properties of the visual stimulus results in retinotopic maps, i.e., visual field representations in the brain. In other words, retinotopic maps are estimates of the spatial preference of cortical surface vertices with respect to different locations of the visual field. The polar coordinate system is the preferred coordinate system for reporting the spatial location of visual stimuli in the visual field in visual neuroscience. Hence, polar angle maps are retinotopic maps reflecting the polar angle in the visual field to which a vertex is most responsive. On the other hand, eccentricity maps reflect the distance from the center of the visual field (i.e., the fixation point) to which a vertex is most responsive. Note that our work here focused on polar angle maps from the left hemisphere, in which unique variations in the dorsal portion of early visual cortex are seen. For more details of retinotopic mapping data analysis, please refer to Benson et al. (2018).

	\subsection{Region of interest}
	Similar to our previous study\cite{Ribeiro2021}, we narrow our analysis to visual areas defined by a surface-based probabilistic atlas\cite{Wang2015}. This probabilistic atlas includes the dorsal and ventral portions of V1/V2/V3 (not including the fovea) and 19 higher-order visual areas. We slightly modified the atlas by extending V1/V2/V3 ROIs to include their foveal confluence and combining ventral and dorsal components. The final region of interest (ROI) used for model training was drawn to include V1/V2/V3 and their foveal confluence, and 18 higher-order visual areas as defined by Wang et al. (2015), without any discontinuity between visual areas. After vertex selection, the remaining surface consisted of 3,267 vertices for the left hemisphere, with 18,998 edges between pairs of vertices.
	
	\subsection{Data for geometric deep learning}
	In our modality transfer task, we trained a geometric deep learning model (see the following section) to predict retinotopic polar angle maps from curvature and myelin maps in a vertex-wise manner. Hence, the input surface was that of our region of interest (as abovementioned) selected from the HCP 32k fs\_LR standard midthickness surface space (\verb|S1200_7T_Retinotopy181.L(R).|\-\verb|sphere|.\verb|32k_fs_LR.surf.gii|). Specifically, the input’s surface topology (connectivity among the ROI surface vertices and spatial disposition) was the same for all participants. Additionally, curvature and myelin maps served as vertices features (Figure \ref{fig1}A), which differed across individuals. Further, the output for the model, or task goal, were polar angle maps comprising vertex-wise polar angle values. The left hemisphere represents the right visual field, which in polar angle includes 0$^{\circ}$-90$^{\circ}$ (upper right visual field) and 270$^{\circ}$-360$^{\circ}$ (lower right visual field), having a discontinuity at the horizontal meridian. To avoid issues with this discontinuity, we shifted the polar angle values of ground truth maps so that the left hemisphere was trained with the point of wrap-around (from 360$^{\circ}$ to 0$^{\circ}$) positioned at the horizontal meridian in the contralateral hemifield. After training, predicted and ground truth polar angle maps were shifted back to their original range.
	
	The original HCP 7T Retinotopy dataset consists of functional and structural data from 181 participants. For our analysis, we randomly split participants into three datasets: training (161 participants), development (10 participants), and test (10 participants). Note that we kept the same splits as in our previous study, in which we report participants’ IDs in both the development and test sets. The training set was used for model training to learn the correspondence between the polar angle maps and the anatomical maps. Although we do not perform hyperparameter tuning here, the development set was used for an unbiased selection of the best-performing model. Finally, the selected model’s performance was evaluated by assessing the predicted maps for each participant in the test dataset.

	\subsection{Model architecture and training}
	We implemented our previously validated model architecture for this modality transfer task. This architecture consists of 12 spline-based convolutional layers\cite{Fey2018}, with the number of feature maps increasing to a maximum of 32 feature maps per layer and then decreasing back to one final feature map. Spline-based convolution is a spatial filtering approach used to aggregate information locally around a vertex’s neighborhood. These filters exploit information given by relative positions of the neighbor vertices to the reference vertex in addition to the information encoded in the connectivity, edge weights, and vertex features. Thus, SplineCNN aggregates vertex features in local neighborhoods weighted by learnable parameters of a continuous kernel function\cite{Fey2018}, enabling a new vertex representation that serves as input to the next convolution layer. In other words, spline-based convolution kernels learn vertex representations by iteratively transforming and aggregating feature vectors of their neighboring vertices. Previously, Fey et al. (2018) found that the relative distance between vertices computed in Cartesian coordinates and the degree of B-spline basis m = 1 resulted in better performance than using a higher degree of B-spline basis or spherical coordinates. Therefore, we used these same parameters and fixed the kernel size (k) to 25 for all layers. Finally, convolutional layers were interleaved with nonlinear activation functions - the exponential linear unit (ELU), batch normalization, and dropout (p = 0.10). 
	
	Model training was carried out for 200 epochs with stochastic gradient descent (batch size = 1) and a learning rate at 0.01 for 100 epochs, then adjusted to 0.005, using Adam optimizer. For this modality transfer task, the learning objective was to reduce the difference between predicted polar angle maps and ground truth in a vertex-wise manner, expressed by the smooth L1 loss function (Equation \ref{equation:loss}).
	
	\begin{equation}
	loss(\widehat{y}^{*}, y^{*}) = \frac{1}{n}\sum_{i}z_{i}
	\label{equation:loss}
	\end{equation}
	
	where \(y^{*}\) and \(\widehat{y}^{*}\) are, respectively, the ground truth and predicted value weighted by individual-specific explained variance \(R^{2}\) from the pRF modeling procedure\cite{Benson2018b} (an indirect measure of ground truth reliability), \textit{n} is the total number of vertices, and \(z_{i}\) is given by:
	
	\begin{equation}
	z_{i} = \left\{\begin{matrix}
	0.5({\widehat{y}^{*}}_{i} - {y^{*}}_{i})^{2}, if |\widehat{y}^{*}_{i} - {y^{*}}_{i}| < 1
	\\ 
	|\widehat{y}^{*}_{i} - {y^{*}}_{i}| - 0.5, otherwise
	\end{matrix}\right. 
	\label{equation:loss2}
	\end{equation}
	
	Models were implemented using Python 3.7.10, Pytorch 1.6.0, and PyTorch Geometric\cite{Fey2019} 1.6.3. In addition, surface plots were generated using Nilearn, a Python module for fast statistical learning on neuroimaging data\cite{Abraham2014}. The training was performed on a high-performance computing cluster using NVIDIA Tesla V100 Accelerator units.
	
	\subsection{Model selection and evaluation metric}
	We trained five models with the same hyperparameters (as previously described) but different learnable parameters due to the random initialization. Following that, we selected the best performing model among those five by determining the difference between predicted and ground truth polar angle values in the development dataset, estimated by the smallest difference between two angles, given by:
	
	\begin{equation}
	MIN(|\widehat{\theta} - \theta|, |\widehat{\theta} - \theta + 2\pi|, |\widehat{\theta} - \theta - 2\pi|)
	\label{equation:chap2eq3}
	\end{equation}
	
	for \(0 < \theta < 2\pi\). We then averaged the vertex-wise difference across participants and within the combined V1/V2/V3 visual areas. The minimum difference between two angles was similarly used to determine the difference between the original prediction (with intact input features) and the altered one (with perturbed input features).
	
	\subsection{Perturbation-based explainability}
	Our perturbation-based approach artificially occludes the features associated with a set of vertices and monitors the difference in polar angle predictions (intact vs. modified input). Specifically, we were interested in determining where and which features and vertices were relevant to individual variability seen in the dorsal portion of early visual cortex, where polar angle maps have shown to be highly variable across individuals\cite{Allen2021,Arcaro2015a,Benson2018a,VanEssen2018}. To answer our first line of inquiry, i.e., defining the \textbf{spatial location} of important vertices, we iteratively selected a target vertex, determined its N-hop neighborhood, replaced their features (myelin and curvature) with constants, and generated new polar angle maps. Constants were the mean curvature (0.027) and mean myelin values (1.439) across vertices and participants in the training set. Thus, we systematically occluded both curvature and myelin maps by sweeping a fixed-value (non-informative) kernel over the input surface. After features were modified (or occluded), new predictions were generated for all participants in the test dataset with the previously trained model. Hence, this process results in V (number of target vertices) × 10 (test dataset size) new predictions given a fixed neighborhood size (N). To reduce the search space of target vertices, we only consider those from V1/V2/V3, totaling 1098 vertices. Finally, for each target vertex, the difference between polar angle predictions with intact versus modified input, the induced error, was determined in a vertex-wise manner and averaged across vertices in the dorsal portion of early visual cortex (Figure \ref{fig1}B) and participants in the test dataset, resulting in a single value per target vertex representing the saliency score. Therefore, we generated a saliency map comprising saliency scores for each vertex as a function of their spatial location on the surface. 
	
	The neighborhood size can be adjusted such that it can include only the target vertex (N = 0); target vertex and vertices directly connected to it (N = 1); target vertex, vertices connected to the target vertex and their new connections (N = 2); and so forth. Varying neighborhood size can answer how sensitive our model is to changes in very localized (N = 0) or more extensive areas covering input features. Thus, we also tested the effect of different neighborhood sizes on saliency scores for the vertices in the dorsal portion of early visual cortex.
	
	Finally, to answer our second line of inquiry, which feature was the most \textbf{important feature} for individual variability, we occluded one feature map at a time instead of both. In other words, we iteratively selected a target vertex, determined its N-hop neighborhood, and replaced intact curvature (or myelin) values with a constant. Then, we similarly generated new polar angle maps for the modified input and computed the difference between predictions with intact versus modified features. Finally, we averaged saliency scores across vertices in the dorsal portion of early visual cortex.

	\subsection{Validation}
	Often overlooked, reliability and validity are desirable qualities for any explainability approach\cite{Leavitt2020}. Reliability is associated with the consistency of the measurement, in our case saliency scores, across different trained models. Conversely, validity refers to the very ability of the method to detect what is meant to be detected, i.e., the importance or saliency of vertices and features. For \textbf{reliability}, we determined the degree of consistency of saliency maps by determining Pearson’s correlation scores between pairs of vectorized saliency maps from all five pre-trained models. To assess the \textbf{validity} of our perturbation-based method, we performed two distinct experiments: (\textbf{1}) we trained new models with relevant features replaced with a constant value, and (\textbf{2}) masked important vertices from the loss function for semi-supervised training. In (\textbf{1}), we occluded features (replaced intact values with constants) from all vertices in the dorsal portion of early visual cortex. We hypothesized that if local information was indeed important and necessary, a model trained with localized features occluded would be unable to learn the unique and accurate representations of polar angle maps seen in the same location. In (\textbf{2}), we masked the same vertices from the first experiment from the loss function for semi-supervised training. Similarly, we hypothesized that in the absence of the prediction goal (polar angle value) during training, a model relying on local information would not learn accurate representations. Although similar to (\textbf{1}), we expected more inaccurate maps in (\textbf{2}), since the loss function would not access the prediction goal for the masked vertices, i.e., models' weights would not be optimized to predict the polar angle values of those vertices.   
	
	\subsection{Data and code availability}
	The data used in this study is publicly available at BALSA (https://balsa.wustl.edu/study/show/9Zkk). In addition, all pre-trained models are available on the Open Science Framework (https://osf.io/f4dez/), and all accompanying Python source code is available on GitHub (https://github.com/Puckett-Lab/explainability\_CorticalSurfaceGDL). Moreover, an executable code notebook hosted on Google Collab is available on GitHub, allowing anyone to load the trained models and generate perturbation-based saliency maps. 
	
	\section{Results}
	
	\subsection{Localizing important vertices: is individual variability determined by local or broad spatial dependencies?}
\begin{figure}[htp]
		\centerline{\includegraphics[width=\columnwidth]{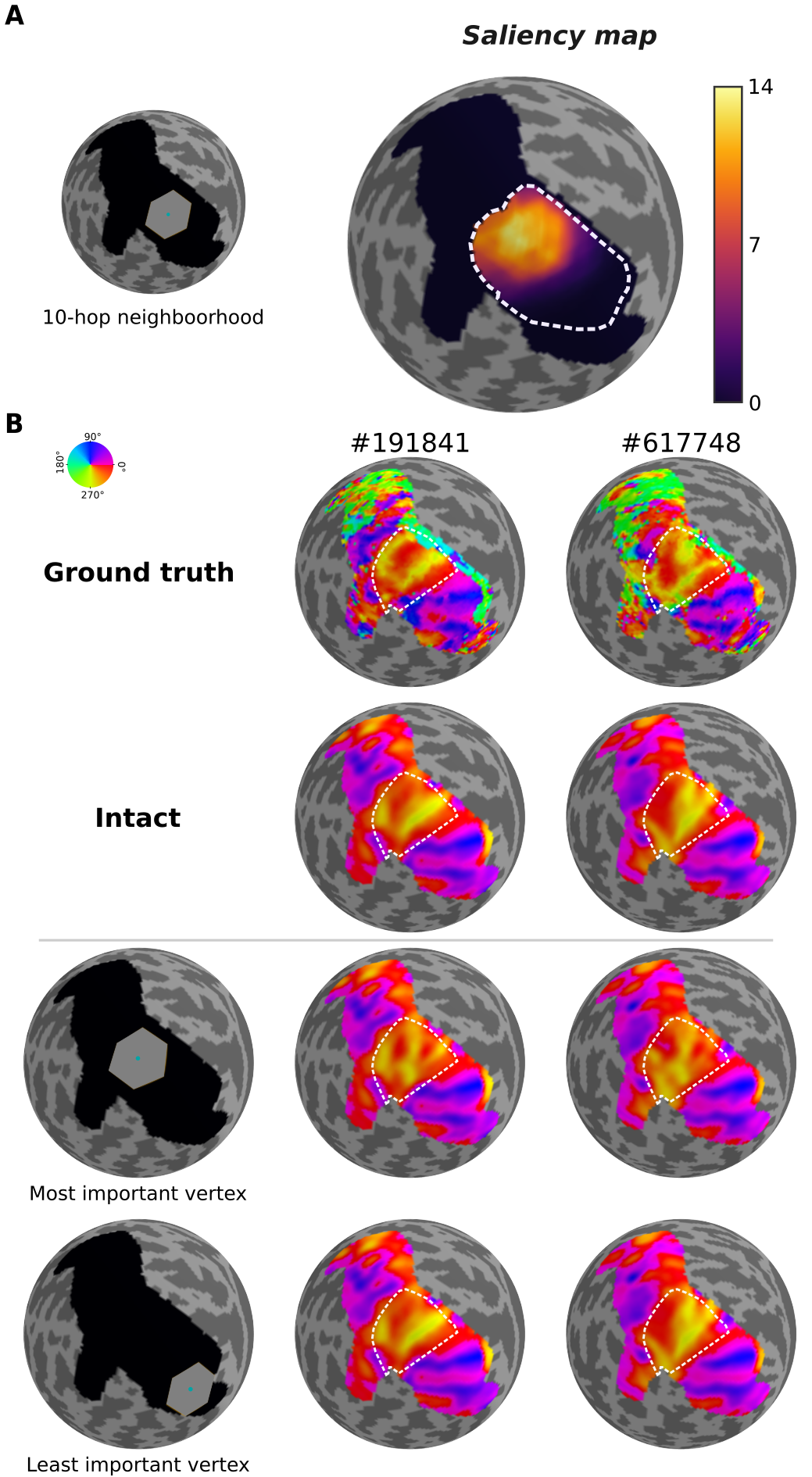}}
		\caption{\textbf{Spatial location of important vertices. A,} Saliency map (right) generated using our perturbation-based explainability approach with a fixed neighborhood size equal to 10-hop (left). \textbf{B,} Upper row shows ground truth polar angle maps from two participants in the test dataset (see HCP ID\#). The second row shows predicted polar angle maps from the same participants using intact input features. The third row shows the location of the most important vertex (left) with its 10-hop neighborhood and predicted polar angle maps when replacing those vertices’ features with constants. Note that, by occluding the most important vertex and its neighborhood, predictions at the dorsal portion of early visual cortex (dashed white line) are less accurate, which relates to the vertex’s high saliency score. On the other hand, the fourth row shows the location of the least important vertex with its 10-hop neighborhood and predicted polar angle maps when replacing those vertices’ features with constants, which does not affect predictions in the dorsal early visual cortex – hence, the lowest saliency score.}
		\label{fig2}
	\end{figure}

	Our first motivational question was where the key vertices underlying individual variability were located. To determine the spatial location of important vertices, we iteratively selected a target vertex, determined its 10-hop neighborhood, replaced their features with constants, and generated new polar angle maps (Figure \ref{fig1}B). Then, we quantified the difference between predictions with intact and modified input and averaged across vertices in the dorsal portion of early visual cortex (Figure \ref{fig1}B) as well as across participants in the test dataset, resulting in a single value per target vertex representing the saliency score. Figure \ref{fig2}A shows the final saliency map, i.e., saliency scores as a function of the spatial location of target vertices on a spherical surface model. The saliency score expresses how much a prediction changes after occluding a specific region in the input; thus, a greater saliency score is equal to a greater change in the model’s prediction. Therefore, the saliency score is a proxy of the importance of a given vertex (and its neighborhood) in generating the most accurate prediction. 
	
	We found that vertices located more dorsally in early visual cortex were the most important for determining appropriate predictions at the dorsal portion of early visual cortex. That is, to predict the most accurate representations of polar angle maps in the specified location, our surface-based deep learning model relies on colocalized input information. This finding can be better appreciated by inspecting the modified predictions from the most and the least (Figure \ref{fig2}B) important vertices. While the occlusion of the most important vertex and its neighborhood (Figure \ref{fig2}B) resulted in less accurate maps at the dorsal portion of early visual cortex (dashed white line), occluding the least important vertex had a minor effect on polar angle maps.
	
\begin{figure}[htp]
		\centerline{\includegraphics[width=\columnwidth]{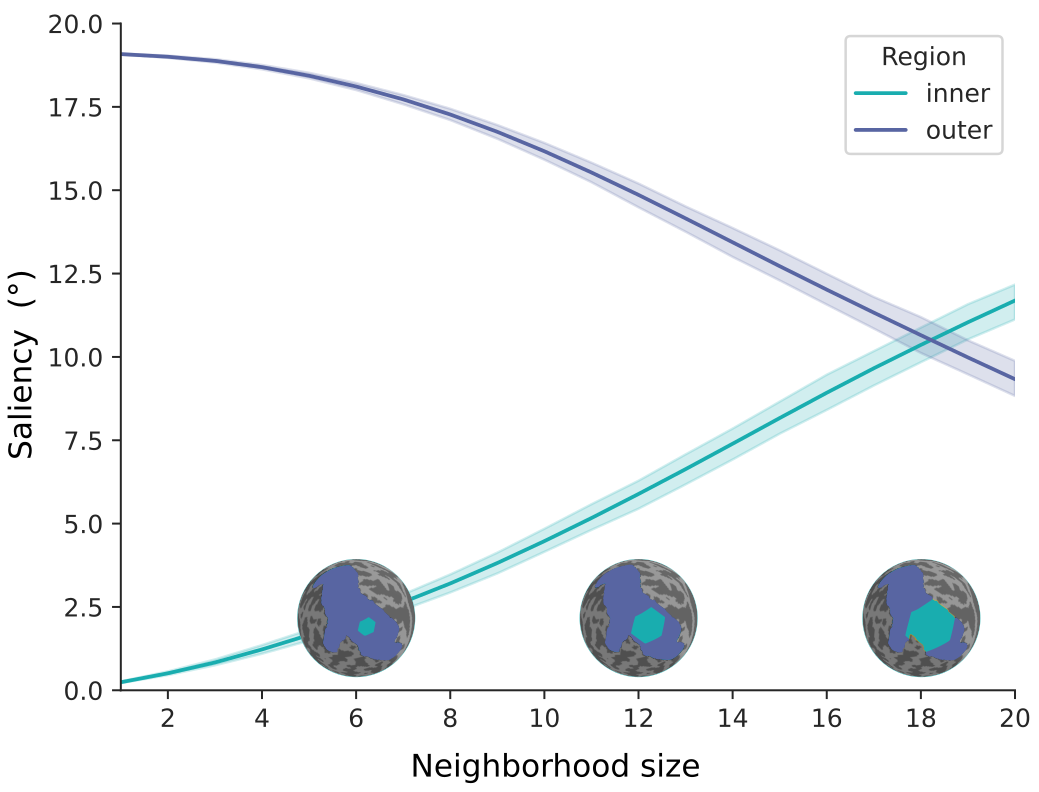}}
		
		\caption{\textbf{Saliency scores as a function of neighborhood size.} The average saliency score (and 95\% confidence interval) of vertices within the dorsal portion of early visual cortex as a function of the size of occluded areas (inner and outer). Spherical surfaces illustrate the inner and outer regions for three neighborhood sizes (6, 12 and 18).}
		\label{fig3}
	\end{figure}

	We also investigated the effect of neighborhood size on saliency scores of vertices in the dorsal early visual cortex. Figure \ref{fig3} shows the average saliency score as a function of neighborhood size (inner region), demonstrating that as neighborhood size increases, so do saliency scores. That is, when occluding a small number of vertices within our region of interest (dorsal early visual cortex), the error is low. However, as more vertices are occluded, the degree of disruption seen across the maps increases. We also show the average saliency scores when we occlude all the vertices outside the neighborhood (outer region) while leaving those within the neighborhood intact. Noteworthy, the larger the neighborhood, the smaller the outer region. As seen with occluding the neighborhood, the saliency scores appear to track with the size of the occluded area in the outer region. Note that when the neighborhood size is over 18, occluding the inner region begins to result in higher saliency scores than those from occluding the outer region. Importantly, beyond this point, the neighborhood begins to cover the totality of the dorsal portion of early visual cortex. Hence, when the inner region is occluded, all the vertices in our region of interest are occluded; when the outer region is occluded, all the vertices in our region of interest are available, providing further evidence that our model is learning local polar angle representations from colocalized input features.

	\subsection{Establishing importance of features: which feature is the most important for individual variability?}
	
\begin{figure}[htp]
		\centerline{\includegraphics[width=\columnwidth]{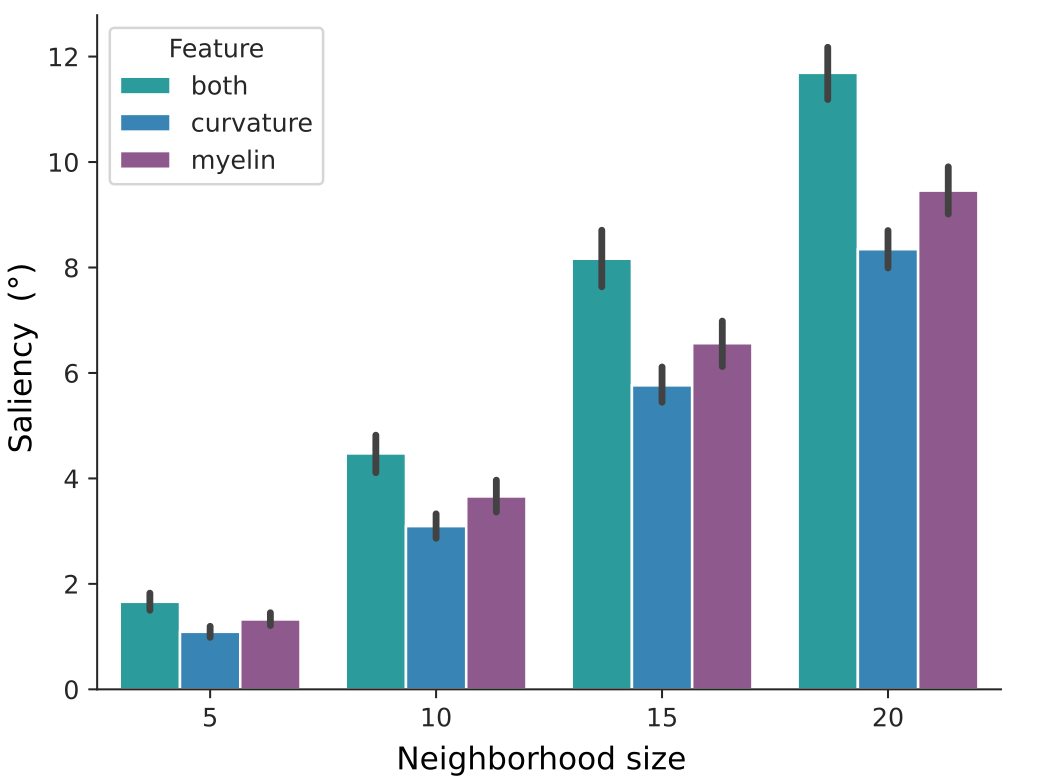}}
		
		\caption{\textbf{Unraveling the importance of input feature maps.} The average saliency score (and 95\% confidence interval) of vertices within the dorsal portion of early visual cortex as a function of neighborhood size and occluded feature map.}
		\label{fig4}
	\end{figure}

	Our next motivational question was which feature was the most important to individual variability, given that unique polar angle representations colocalize with important vertices (Figure \ref{fig2}). To find out which feature map was the most important, we similarly occluded patches of the input maps, but we occluded one feature map at a time. In other words, we iteratively selected a target vertex, determined its neighborhood, replaced the intact curvature (or myelin) values with a constant, and generated new polar angle maps. Then, after we calculated the difference between predictions with intact and modified input, we averaged the saliency score across vertices in the dorsal portion of early visual cortex. Figure \ref{fig4} shows the average saliency score as a function of neighborhood size and feature map. While occluding both feature maps led to the highest saliency scores, we also found that occluding myelin features alone resulted in saliency scores higher than when curvature features were occluded. Additionally, average saliency scores increased as a function of neighborhood size for all features. Therefore, these results suggest that both features are providing useful information to predict retinotopy in the local area, i.e., neither feature is redundant. Nonetheless, our model seems to be differentially relying on myelin feature maps to determine individual variability in the dorsal portion of early visual cortex. 
	
	%The small difference seen between occluding myelin vs. curvature, however, could be due to random initialization of the model; that is, different models could differentially rely on either curvature or myelin. 
	
	\subsection{Reliability: are saliency maps consistent?}
	
	Next, we sought to determine whether our perturbation-based explainability method was reliable, i.e., consistently generated similar saliency maps. We determined the degree of spatial consistency of saliency maps by determining Pearson’s correlation scores between pairs of saliency maps from five randomly initialized models. Saliency maps were generated by occluding myelin only, curvature only, and both input feature maps. Table \ref{tab1} shows the mean correlation scores across all combinations of saliency maps from each condition. Overall, saliency maps were highly consistent across randomly initialized models. This finding shows that regardless of the initial model’s parameters, the models consistently converge to a similar mapping function after training, in which the same vertices play similar roles in determining individual variability.
	
	\begin{table}[h]
		\begin{tabularx}{\columnwidth}{lc}
			\hline
			\textbf{Feature} & \textbf{Mean correlation score ± SD} \\ 
			\hline
			Both             & 0.987 ± 0.007  \\
			Myelin           & 0.988 ± 0.009  \\
			Curvature        & 0.989 ± 0.008  \\ 
			\hline
		\end{tabularx}
		\caption{\textbf{Pearson's correlation scores between pairs of saliency maps from five different models.}}
		\label{tab1}
	\end{table}

	\subsection{Validity: is the method capable of detecting important vertices and features?}
	
\begin{figure}[htp]
		\centerline{\includegraphics[width=\columnwidth]{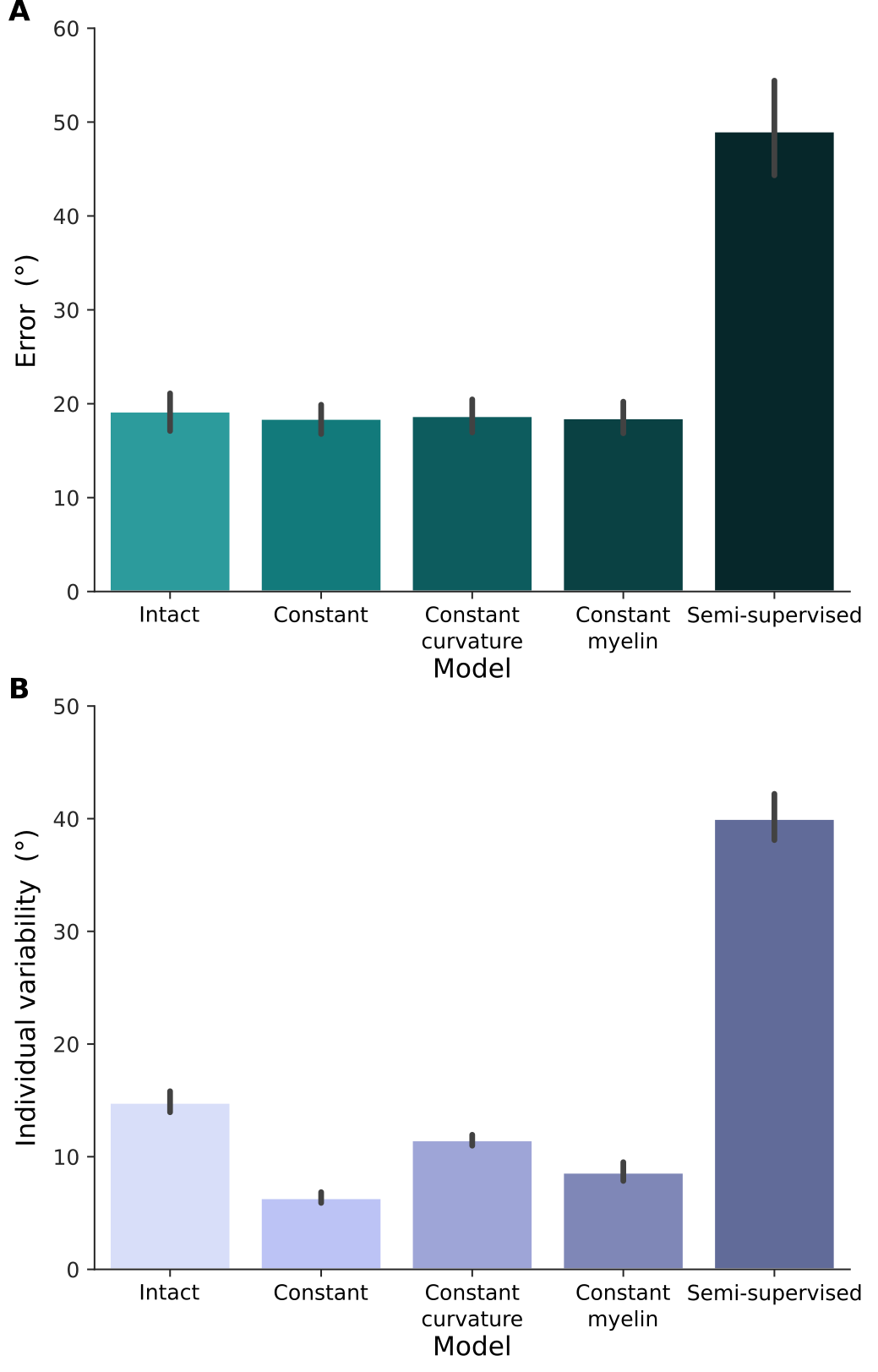}}
		\caption{\textbf{Prediction error and individual variability of models trained on different conditions. A,} Prediction error (difference between ground truth and predicted maps) was averaged across vertices in the dorsal portion of early visual cortex and individuals in the test dataset. The results are from the best model among five randomly initialized models trained for each condition, which was selected based on prediction error in the development dataset. \textbf{B,} Individual variability was defined as the difference between a particular predicted map and each of the other predicted maps in the test dataset in a vertex-wise manner and averaged across all comparisons.}
		\label{fig5}
	\end{figure}

	Finally, to determine the validity of our explainability approach, we performed two distinct experiments: (\textbf{1}) we trained new models with relevant features replaced with a constant value, and (\textbf{2}) masked important vertices out from the loss function for semi-supervised training. A valid explainability method should detect what is meant to be detected, i.e., the importance or saliency of vertices and features. Therefore, we hypothesized that a model relying on local information would not learn accurate representations of individual variability without the important nodes and features previously identified during training. Figure \ref{fig5}A shows prediction errors (difference between predicted and ground truth maps) from models trained on: (\textit{1}) intact features (‘intact’), (\textit{2}) both input features replaced with constants (‘constant’), (\textit{3}) intact myelin but constant curvature (‘constant curvature’), (\textit{4}) intact curvature but constant myelin (‘constant myelin’), and (\textit{5}) masked vertices (‘semi-supervised’). Although replacing features from important nodes by a constant has resulted in models with similar accuracy to the one trained on intact features, completely masking important vertices from the loss function resulted in a much more inaccurate model. 
	
	Given that the overall prediction error was similar across models trained on modified and intact features, we also determined the overall level of individual variability (Figure \ref{fig5}B). To do so, we computed the difference between a particular predicted map and each of the other predicted maps in the test dataset in a vertex-wise manner and averaged across all comparisons\cite{Ribeiro2021}. Our results suggest that replacing both input feature maps with constants led to models that predicted less variable (and hence less appropriate) polar angle maps, emphasizing neither feature is redundant. Furthermore, the model trained on at least intact myelin features (constant curvature) generated more idiosyncratic predictions than the one trained on intact curvature (constant myelin), but the difference is small. Overall, these results suggest that our perturbation-based explainability method was indeed capable of identifying important vertices and features that were underlying individual variability in predicted maps.
	
	\section{Discussion}
	Here, we proposed a perturbation-based explainability method for surface-based deep learning. We showed that our method can detect the spatial location of salient vertices and determine which feature was the most relevant for individual variability in the predicted maps. Briefly, we found that our cortical surface-based deep learning model relies on colocalized input features to predict unique variations in retinotopic (polar angle) maps. Our results suggest that both anatomical features (curvature and myelin patterns) provide useful information to predict the retinotopic organization in the dorsal portion of early visual cortex. Notably, the model trained on intact features was found to differentially rely on myelin feature maps to predict individual variability. Our validation experiments supported this finding, in which the model trained only with intact myelin (constant curvature) generated more idiosyncratic predictions than the one trained with only intact curvature (constant myelin). We caution, however, that the small difference seen between curvature vs. myelin could be due to random initialization of models. Therefore, a more comprehensive study is required to delineate the differential roles of anatomical features to retinotopy. Finally, our reliability experiments suggest that our explainability method consistently detects important vertices. We found that saliency maps were highly consistent across five randomly initialized models, indicating that models repeatedly converge to a similar mapping function after training, in which the same vertices play similar roles in determining individual variability. In summary, our findings suggest that colocalized anatomical features underlie, at least in large part, the individual variability often seen in the dorsal portion of human early visual cortex.
	
	Anecdotal evidence from several studies has suggested that the dorsal portion of early visual cortex is more variable across people than its ventral counterpart, particularly dorsal V3\cite{Allen2021,Arcaro2015a,Benson2018a,VanEssen2018}. Similarly, research suggests that there may be a more complicated organization of an analogous area in non-human primates\cite{Angelucci2015,Gattass1988,Zhu2019}. While the research field conventionally assumes that dorsal V3 is a single visual area in the human brain, the same assumption is under continuous debate for non-human primates. More recently, Yu et al. (2020) have found that the unusual inhomogeneities in polar angle maps of marmoset monkeys in an analogous area to human dorsal V3 may emerge from differing smoothness (topographic continuity within a visual area) and congruence (topographic continuity with neighboring visual areas) constraints. These effects were suggested to arise from a combination of colocalized factors such as receptive field overlap, the density of the columnar structure, the extent of intrinsic connections, and the wiring cost of the underlying circuits\cite{Yu2020}. With our perturbation-based approach, we similarly found that our geometric deep learning model for retinotopic mapping exploits colocalized input features to predict unique variations in polar angle maps. As such, extending the previous geometric deep learning model to include other input features and subsequently investigating their contributions using our perturbation-based explainability approach may offer a powerful way forward in elucidating the factors affecting the topographic organization of the visual cortex, as well as how they contribute to the unusual maps seen in human dorsal V3 and analogous areas in non-human primates.
	
	Our perturbation-based explainability framework was developed to serve as a step towards a more transparent deep learning system for surface-based deep learning. Accordingly, reliability and validity were desirable qualities. We assessed the reliability of the proposed framework by determining the degree of consistency of saliency maps from five randomly initialized models. Future work could evaluate the consistency of saliency maps generated with different explainability methods adapted to modality transfer tasks. If different methods converge to a similar set of salient vertices and features, then a more robust conclusion can be drawn about the inner mechanisms of a deep learning model. On the other hand, if salient vertices and features differ, one could perform validity experiments to assess whether different methods are more valid than others, i.e., whether one explainability approach is better suitable to detect important vertices and features than others. Finally, since our conceptual framework is not limited to this case study, we provide all source code in the hope it serves as a starting point for future work in explainable AI.
	%Finally, given that our conceptual framework is not limited to this case study, we provide all source code in the hope it serves as a starting point for future work in surface-based explainable AI.

\end{document}